# DEVELOPMENT OF A DUAL SENSOR HEAT CONTROL SYSTEM

Adamu Murtala Zungeru[1] and Mahmud Shehu Ahmed[2]

[1]School of Electrical and Electronic Engineering, University of Nottingham, Jalan Broga, 43500 Semenyih, Selangor Darul Ehsan, Malaysia
adamuzungeru@ieee.org
[2]Department of Electrical and Electronic Engineering, Federal University of Technology, Minna, Nigeria
mahmudmsa@yahoo.com

*ABSTRACT*

*Convenience and safeguarding our home appliances have become an important issue when dealing with an advancement and growth of an economy. This research focuses on the design and construction of a Dual Sensor heat-monitoring system. The circuit works by monitoring temperature from an external input and comparing the temperature level with that of a preset temperature value. The power output of the circuit is cut off or switched OFF or an alarm is triggered ON if the temperature of the external input is equal to or, greater than the preset temperature value. The methodology involves the application of linear precision temperature sensors i.e., they generate a voltage that is directly proportional to the temperature. Basically the system is constructed using temperature sensors and comparators. The system is powered using a 12V power supply. The results of the tests showed that the power output of the circuit is switched OFF hence switching OFF the heating device or an alarm is triggered ON when the device exceeded a preset temperature level. The general operation of the system and performance is dependent on the temperature difference between the preset temperature value and external temperature intended to be monitored. The overall system was tested and found perfectly functional.*

## 1. INTRODUCTION

Temperature is most often measured environmental quantities which correspond to primary sensations-hotness and coldness. This is due to the fact that most biological, chemical, electronic, mechanical and physical systems are affected by temperature. In many instances, some processes perform better within a range of temperatures. Also, certain chemical reactions, biological processes and even electronic circuits do better within limited temperature ranges. When the needs to optimize these processes arose, the systems used for controlling the temperature within a defined limits are then needed. The temperature sensors are often used in providing inputs to those control systems. However, in the case of too much exposure of some electronic components to high extreme temperature, there will be an advace effect on them which can lead to he damage of the components. Though, some of the components can even be affected and get damaged by low temperature values. Semiconductor devices as well as LCDs (Liquid Crystal Displays) can be affected and get damaged by temperature extreme. As the temperature threshold get exceeded, an immediate action should be taken so as to prolong the lifetime of the system. In these, temperature sensing helps to improve the reliability as well as the lifetime of the system.

Most temperature monitoring devices are designed to respond to a particular (critical) temperature level. They are usually incorporated with different kinds of alarms and light indicator units, which are triggered ON at an unacceptable temperature level. These temperature monitoring devices work with temperature sensors normally transducers which generate accurate voltage output that varies linearly with temperature. They are mainly used for monitoring industrial machines, electric boilers, ovens and other heat energy related activities and this can be done by ensuring the temperature sensor and its leads are at the same temperature as the object to be measured. This usually involves making a good mechanical and thermal contact. If the temperature sensor is to be used to measure temperature in liquid, the sensor can be mounted inside a sealed end metal tube and can then be dipped into a bath or screwed into a threaded hole in a tank.

Temperature sensors provide inputs to those control systems. When temperature limits are exceeded, action must be taken to protect the system. In these systems, temperature sensing helps enhance reliability. In modern electronics, more temperature measuring techniques are available. Several temperature sensing techniques are in widespread usage. The most common of these: Thermocouples, Thermistors and Sensor IC's. These temperature sensors ("transducers"), illustrate a nice variety of performance tradeoffs. Temperature range, accuracy, repeatability, conformity to a universal curve, size and price are all involved. The temperature sensor being used in this system is the LM 335 and it has the following features; (1) Directly calibrated in Kelvin, (2) 1°C initial accuracy available, (3) operates from 400u.A to 5mA, (4) less than the 1H dynamic impedance, (5) easily calibrated and (6) low cost. This makes it preferable to the thermistor which was used in previous related research. Although, the thermistor has a wider temperature range than the LM 335, it suffers from self-heating effects, - usually at higher temperatures where their resistances are lower -and fragile, which makes it inappropriate for the research. Apart from the merits of the temperature sensor, the dual nature of the system extends its application which is not so in previous systems designed for heat monitoring. The Complementary Metallic Oxide Semiconductor (CMOS), which provides reasonable performance, is extensively used in the construction of this system. In previous research relating to temperature monitoring, transistor-transistor logic (TTL) integrated circuits were used. The TTL devices are attributed to high power consumption, limited logic functions, narrow power supply, low compatibility and high overall cost while CMOS on the other hand, provides low power consumption, good immunity to external noise, insensitivity to power supply variations, temperature range capabilities -48°C to 52$^C$C [1]. In general, the most important characteristics of CMOS make it the logic of choice. A typical temperature monitoring device possesses both a temperature sensor and control unit that responds to the input. The design of the system involves the triggering of an alarm to indicate that a preset temperature level has been exceeded.

For clarity and neatness of presentation, the article is outline into five (5) sections. The First Section gives a general introduction of a dual sensor heat control system. Work related to the topic of the research is presented in Section Two. In Section Three, we outline the design and implementation procedures. Section Four presents the experimental results and discussion of the results. In Section Five, we conclude the work with some recommendations. Finally, the references are presented at the end of the paper.

## 2. RELATED WORK

Temperature instruments and temperature indicators are designed for temperature monitoring and analysis. These instruments either come equipped with integral temperature sensors, or require temperature sensor inputs. Temperature is the subject of such devices. The concerned sensors are the heart of the devices. The first recorded thermometer was produced by the Italian, Santorio (1561-1636) [2]. As with many inventions the thermometer came about through the work of many scientists and was improved upon by many others. In a related work, Galileo Galilei [3], invented another thermometer. However the instrument he invented was not able to carry the name thermometer, as to become a thermometer an instrument must be able to measure temperature differences. The predecessor to the thermometer, the thermoscope is a thermometer without a scale, that is, it indicates differences in temperature only, that is to say that, it only shows if the temperature is higher, lower or the same, but unlike a thermometer it cannot measure the difference nor can the result be recorded for future reference. To add on, Gabriel Fahrenheit [4] was the first person to make a thermometer using mercury. Fahrenheit used the newly discovered fixed points to devise the first standard temperature scale for his thermometer, and he divided the freezing and boiling points of water into 180 degrees, whereby 32 was the chosen as the reference value (the lower fixed point) as those produced a scale that would not fall below zero even when measuring the lowest possible temperatures that he could produce in his laboratory, a mixture of ice, salt and water. Though also, the Fahrenheit divided his scale into 100 degrees using blood temperature (incorrectly measured) and the freezing point of water as fixed points which is not true. The Fahrenheit scale is still in use today. While others are research done by Anders Celsius (1701-1744) [5], and Sir William Thomson [6]. Any temperature measuring device is usually termed thermometer. Some of the recent work considers different approaches for automation and monitoring different system. In [7,8,9], the authors consider the use of infrared rays to count the number of passengers in a car and also remotely control home appliances via short message services for the purpose of security and human convenience.

Besides, most of the papers mentioned above do not consider cost, reliability and durability in their design procedure, and above all, this paper uses simple and easy to get components to achieve its desired goals.

## 3. SYSTEM DESIGN AND IMPLEMENTATION

This section will discuss the design procedure and the basic theory of components used for this work. The section is further divided into two sub-sections as circuit design analysis and system implementation.

### 3.1. The Circuit Design Analysis

The circuit is designed based on the mode of operation and functions of each component provided by the data sheets. The whole component manufacturers' specifications were carefully put into consideration. The circuit is aimed at minimum or limited number of components for simplicity. The modular design of the system is shown in the block diagram below.

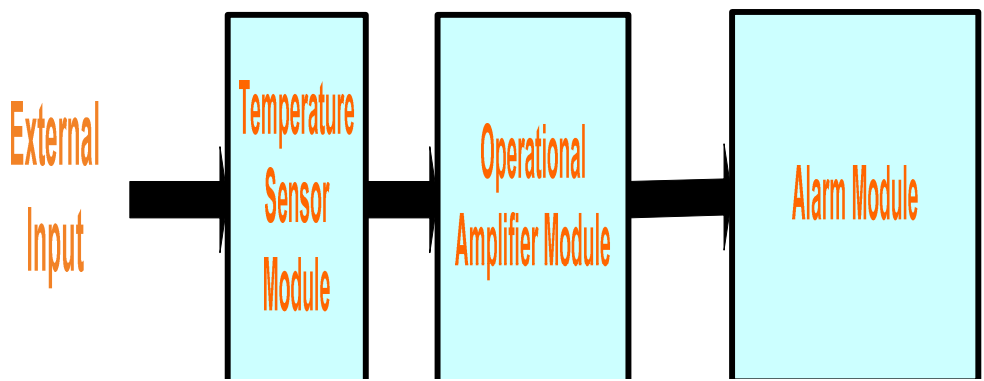

Fig. 1: Block Diagram of the Modular Design of the System

### 3.1.1. Power Supply Unit

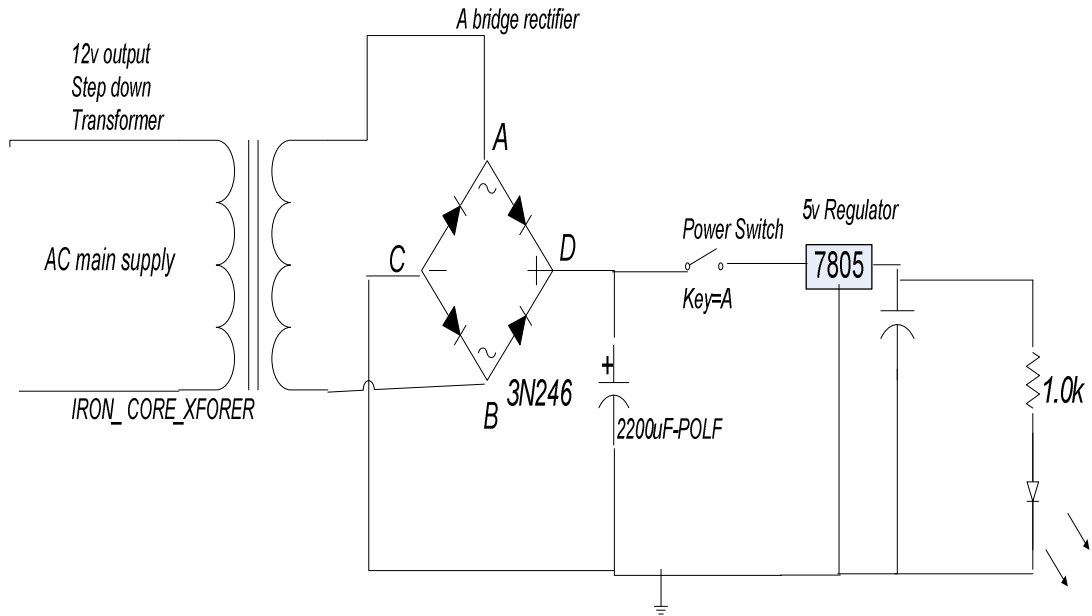

Fig. 2: The power supply module.

The power supply unit as shown in Fig. 2 is designed to provide 5V supply to the circuit. The voltage from an A.C mains supply is stepped down to 12V using a step down transformer and converted to D.C using the bridge system of rectification. This is done by connecting the A.C. supply across one diagonal AB, and D.C output is taken from the other CD line. The bridge rectifier's system output still has a lot of ripples that has to be smoothed out in order to generate genuine D.C. Capacitors resist changes of voltage across them. Hence they are used to provide the desired smoothing action. For 7% ripple, the ripple voltage is calculated thus;

$$Peak\ Voltage, V_{pk} = Vs \times \sqrt{2} \quad (Vs = secondary\ voltage) \tag{1}$$

$V_{pk} = 12V x\ 1.414 - 16.97\ V_{pk}$

$V_{ripple} = V_{pk} \times 7\% = 16.97V\ x\ 0.07 = 1.1879 V_{ripple}$

And the value of the capacitor $C = \frac{I_{load}}{2fV_{ripple}}$

Where I $_{load}$ is 0.2A and frequency from PHCN supply f, = 50Hz.

$$C = \left(\frac{0.2A}{2 \times 50 \times 1.1879}\right) = 1684uF$$

Hence a 2200uF capacitor is used since it's close to 1684uF. A switch is used for opening and closing the complete circuit. A 5V regulator (7805) is incorporated into the circuit to provide regulated 5V supply from the 12V. A light emitting diode is used to indicate the presence of electrical current through the circuit. A resistor is connected in series with the light emitting diode (2.7V) in the forward bias mode to limit electric current. Assuming electric current through the diode is 3.2mA. The value of the series resistor is given below;

$$R = \left(\frac{Voltage\ across\ the\ resistor}{Current\ flowing\ through\ the\ resistor}\right) \quad (2)$$

$$R = \left(\frac{5 - 2.71}{0.0032A}\right) = 718.75 \Omega$$

The 12V and 5V power supplies through the power unit are for the purpose of the integrated circuits and alarm output respectively.

### 3.1.2. The Temperature Unit

Below is the schematic diagram of the temperature sensor unit:

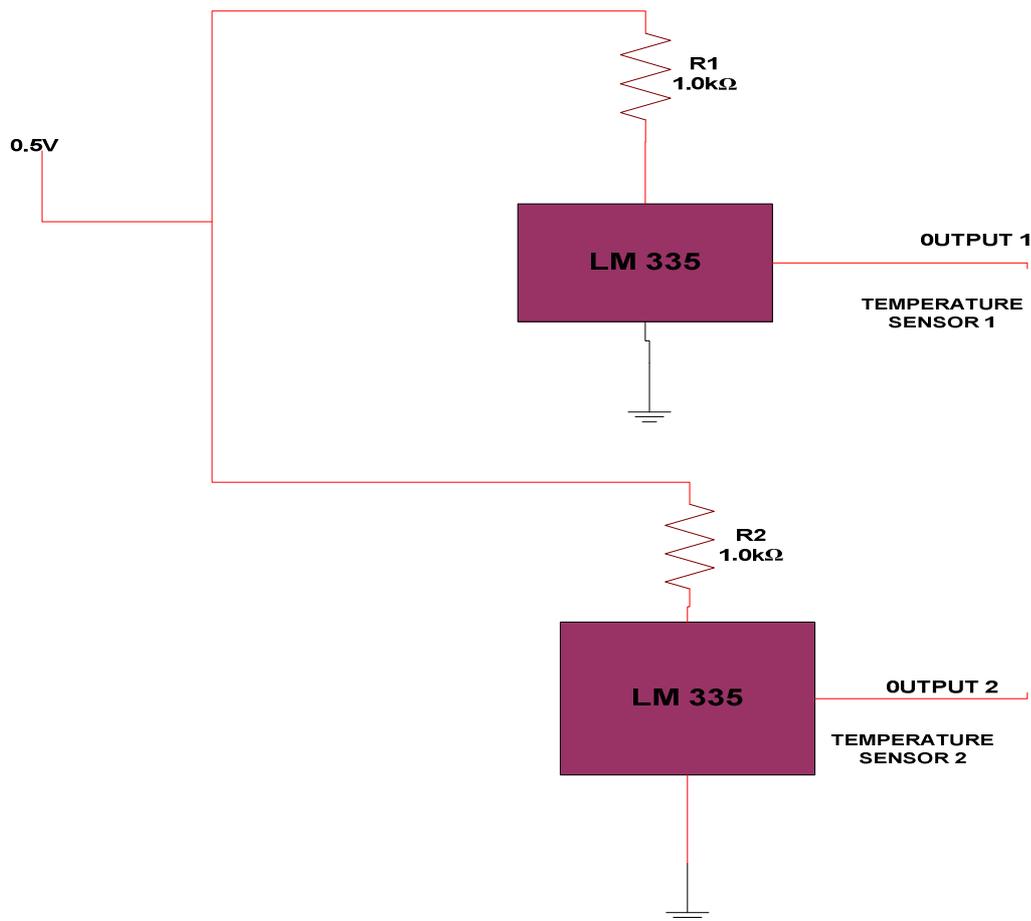

Fig. 3: Circuit diagram of the temperature sensor unit

As shown in Fig. 3 above, the temperature sensor module consists of two LM 335 temperature sensors. Each provides temperature measurement in Kelvin with respect to the voltage. The LM 335 provides a linear relationship between temperature and voltage which is 10mV/°K. The temperature sensors receive input from an external device and they in turn act as input to the rest of the circuit. In other words, the other parts of the circuit respond to the signal from the temperature sensors. The LM 335 measures temperature in Kelvin hence its output voltage is relatively large. For example, at a temperature of 25°C i.e. 298.2K, it has a corresponding voltage of 298.2mV.

### 3.1.3. The Operational Amplifier Unit

Due to the relatively large value of the output voltage of the temperature sensor, the output is connected to an operational amplifier in the subtraction mode. The LM 324 operational amplifier is used to effect subtraction as shown in Fig. 4.

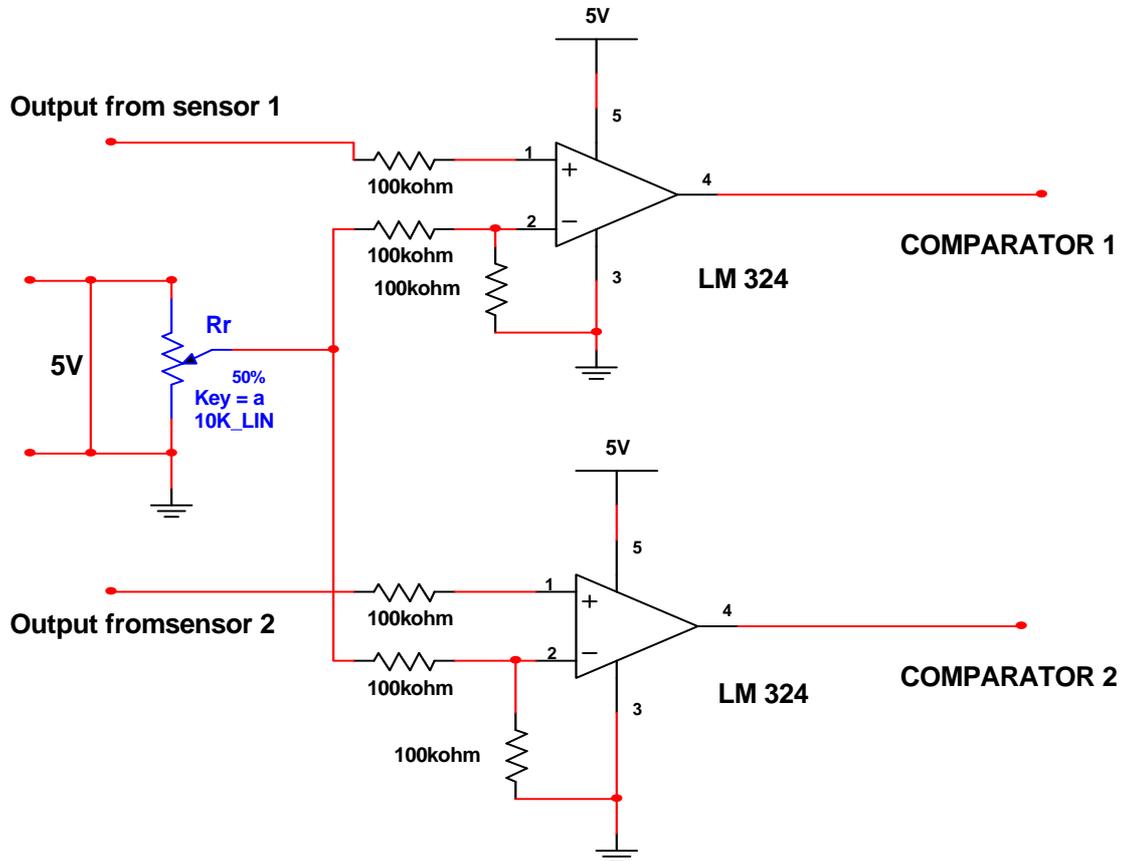

Fig. 4: Circuit diagram of operational amplifier unit

The circuit of Fig. 4 is specified by the component data sheet, RR (10kΩ) is a potential divider for applying 273mV at the negative input of the operational amplifier. The voltage of the positive input of the device is subtracted by 273mV. The output of the LM 324 signifies temperature measurement in degrees Celsius. Two of such operational amplifiers are used in the circuit as shown in Fig. 4.

### 3.1.4. Comparator Unit

The outputs of the operational amplifiers are connected to their corresponding comparators as shown in Fig. 5.

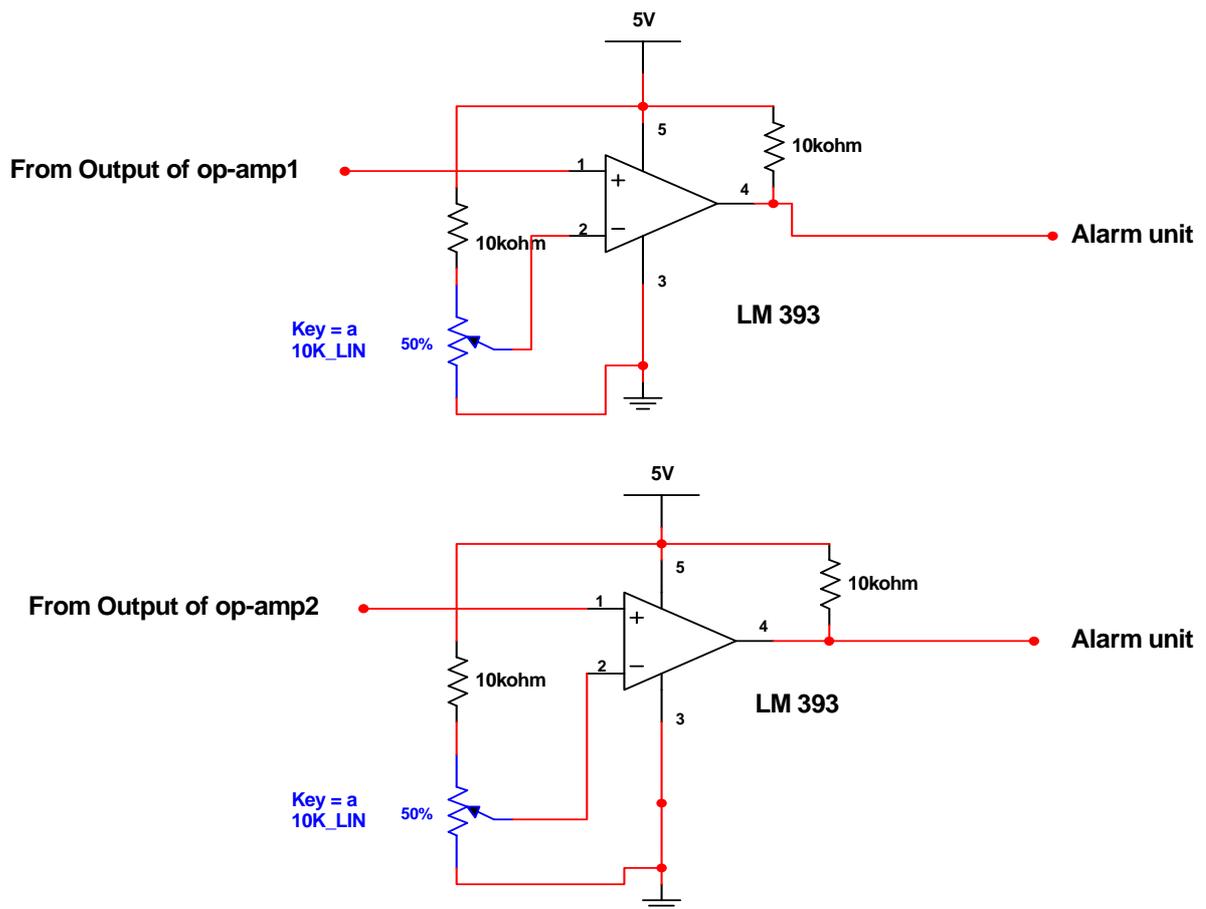

Fig. 5: Circuit diagram of the comparator unit

The comparator module as shown in Fig. 5, involves two comparators (LM 339). The outputs of the corresponding operational amplifiers are connected to the non-inverting input of the comparator. The inverting inputs $V_{in}$ (-) are referenced to a particular voltage through a 10kΩ variable resistor. The resistor circuit that is related to the inverting input $V_{in}$ (-) is used for adjusting a particular circuit's response. Assuming the inverting input of a particular comparator is adjusted to 300mV and the non-inverting input is 200mV. The initial output condition of the corresponding comparator is logic 0 as shown in the relationship below:

$V_{in}(+) > V_{in}(-)$ = logical 1 output

$V_{in}(+) < V_{in}(-)$ = logical 0 output

Hence if the input voltage at the non-inverting pin becomes greater (temperature increases) than that of the voltage at the inverting pin, the output of the comparator changes from logical 0 to logical 1 [1]. By adjusting the variable resistor, different responses are achieved by the circuit.

### 3.1.5. The Alarm Unit

The Tone Generator Unit is built on the 555 Timer IC, but in the Astable mode. Unlike in the Monostable state, the Astable mode has no stable state; the output is continually changing between 'low' and 'high'.

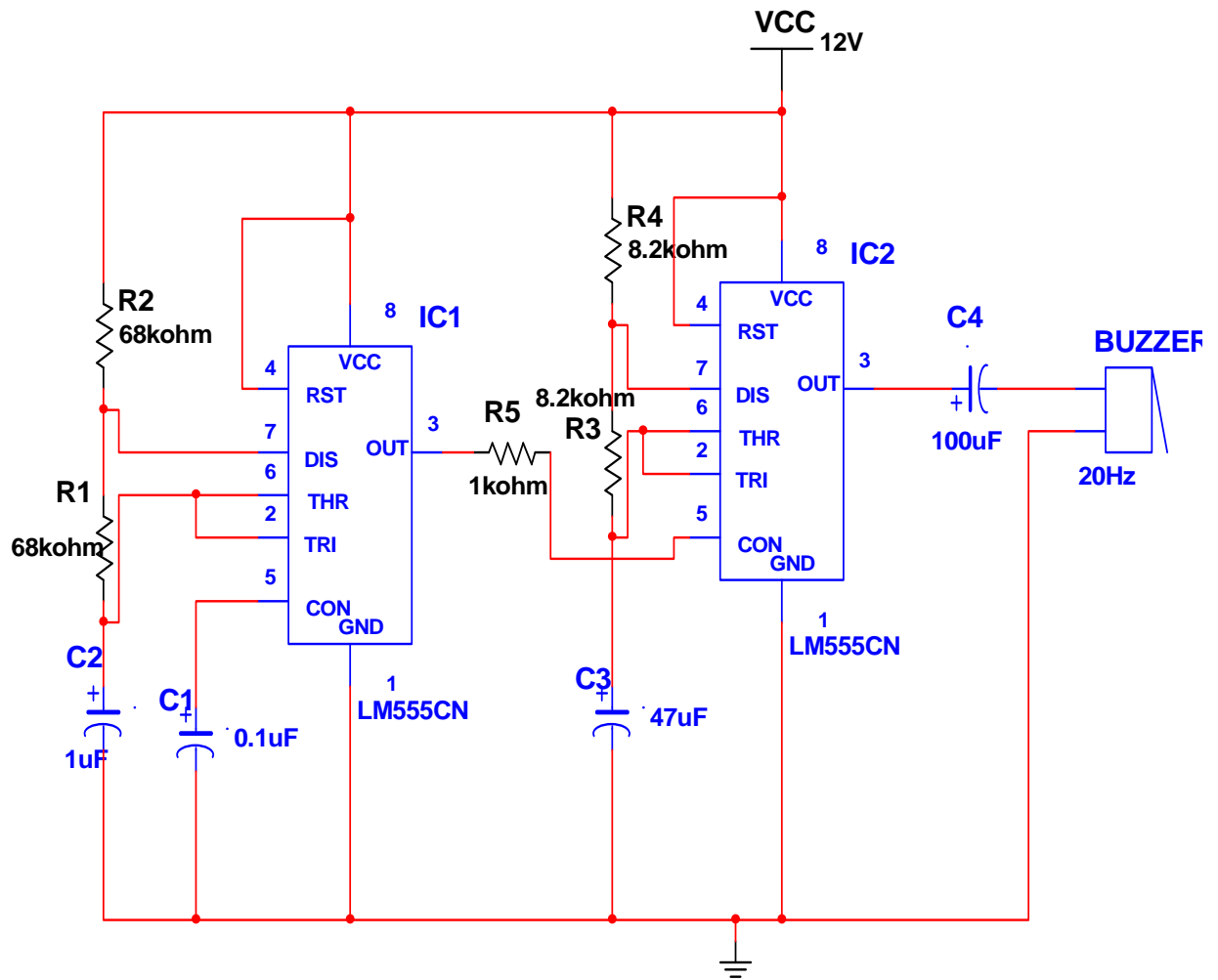

Fig. 6: Circuit diagram of the tone generation unit

### 3.1.5.1. Astable Operation

With the output high (+Vs), the capacitor $C_2$ is charged by the current flowing through $R_1$ and $R_2$. The threshold and trigger inputs monitor the capacitor voltage and when it reaches $2/3 V_S$ (threshold voltage), the output becomes low and the discharge pin (pin7) is connected to 0V. The capacitor now discharges with current flowing through R2 in the discharge pin. When the voltage falls to 1/3 Vs (trigger voltage), the output becomes high again and the discharge pin is disconnected allowing the capacitor to start charging again. This cycle repeats continuously unless the reset input is connected to 0V which forces the output low while, a reset is 0V.

**CALCULATION**

$$T_{ON} = 0.7 \; x \; (R_1 + R_2) \; x \; C_2 \tag{3}$$

$$\text{And } T_{OFF} = 0.7 \text{ x } R_3 \text{ x } C_3 \tag{4}$$

Where, T = time period in seconds (s) = $T_{ON}$ + $T_{OFF}$

With $R_1$ = $R_2$ = 68KΩ and $C_2$ = 1uF

$T_{ON} = 0.7 \times (2 \times 68 \times 10^3) \times 10^{-6}$

$T_{ON} = 0.0952$ seconds.

With R3 = 8.2KΩ and C3 = 47uF, $T_{OFF} = 0.7 \times (8.2 \times 10^3) \times 47 \times 10^{-6}$

$T_{OFF} = 0.26978$ seconds, and $T_{OFF} \approx 0.27$ seconds.

Therefore, T = 0.0952 + 0.27

T = 0.3652 seconds.

Since frequency, F = 1/T, it then implies that F = 1/0.3652

F = 2.74 Hz.

The tone generator unit is controlled by the timing unit via a transistor (BC 108), which acts as a switch to trigger the generator unit ON. Hence, once the timing period is elapsed, the generated tone also ceases.

### 3.1.6. Choice of Components and Analysis

The component selection involves low cost, reliability and availability. The design is also made as simple as possible with the use of a limited number of components. The 4000 series complementary metal oxide semiconductor (CMOS) integrated circuit is intensively used for the design due to its considerable merits over the transistor-transistor logic (TTL) series which is not suitable for this particular design despite the fact that it has high speed which is insignificant to this research. In designing the power unit, three items were taken into consideration: transformer, rectifier and filter. The choice of the transformer used is a step down transformer, so as to step down the input voltage from the Power Holding Company of Nigeria (PHCN) to 12V. "Next is the choice of rectification; there are three types of rectifier systems available for use in power supplies: half-wave, full wave and bridge. The bridge rectifier system was chosen for A.C. to D.C. conversion for the following reasons:

(1) The four diode bridge rectifier provides a greater D.C. output voltage than the center tapped full-wave rectifier circuit.
(2) The need for a center tapped transformer is eliminated.
(3) The PIV is one half that of the center-tap circuit.
(4) The full-wave bridge rectifier utilizes both half-cycles of the input A.C. voltage to produce the D.C. output. Unlike the half-wave rectifier.

But, it is a more expensive system of rectification because, it requires four diodes. The major components of the design are as follows; (1) LM 335 (temperature sensor), (2) LM 393 (comparator), (3) LM 324 (operational amplifier) and (4) NE555 (Oscillator).

### 3.1.6.1. LM335

The LM 335 is a convenient 2-terminal temperature sensor that behaves like a Zener diode with a voltage of +10mV/°K (i.e. It develops an output voltage proportional to absolute temperature). For example, at 25°C (298.2°K), the LM 335 acts like a 2.982 volts Zener. It comes with an initial accuracy as good as ±1°C and it can be externally trimmed. A single point calibration can typically improve its accuracy to ±0.5°C max over a -55°C to 125°C range. These features of the LM 335 are used in the system for precise response to a particular temperature. The LM 335 was chosen for this system because of its linearity, accuracy, features and ease of designing the necessary support circuitry.

### 3.1.6.2. Comparator (LM 393)

The LM393 series consists of two independent precision voltage comparators with an offset voltage specification as low as 2.0mV max for two comparators, which were designed specifically to operate from a single power supply over a wide range of voltages. Operation from split power supplies is also possible and the low power supply current drain is independent of the magnitude of the power supply voltage. These comparators also have a unique characteristic in that the input common-mode voltage range includes ground, even though operated from a single power supply voltage. The comparator has a wide supply (Voltage range: 2.0V to 36V, Single or dual supplies: ±1.0V to ±18V), Very low supply current drain (0.4 mA) independent of supply voltage, input common-mode voltage range includes ground, Differential input voltage range equal to the power supply voltage. Application areas include limit comparators, simple analog to digital converters; pulse, square wave and time delay generators; wide range $V_{CO}$; MOS clock timers; multivibrator and high voltage digital logic gates.

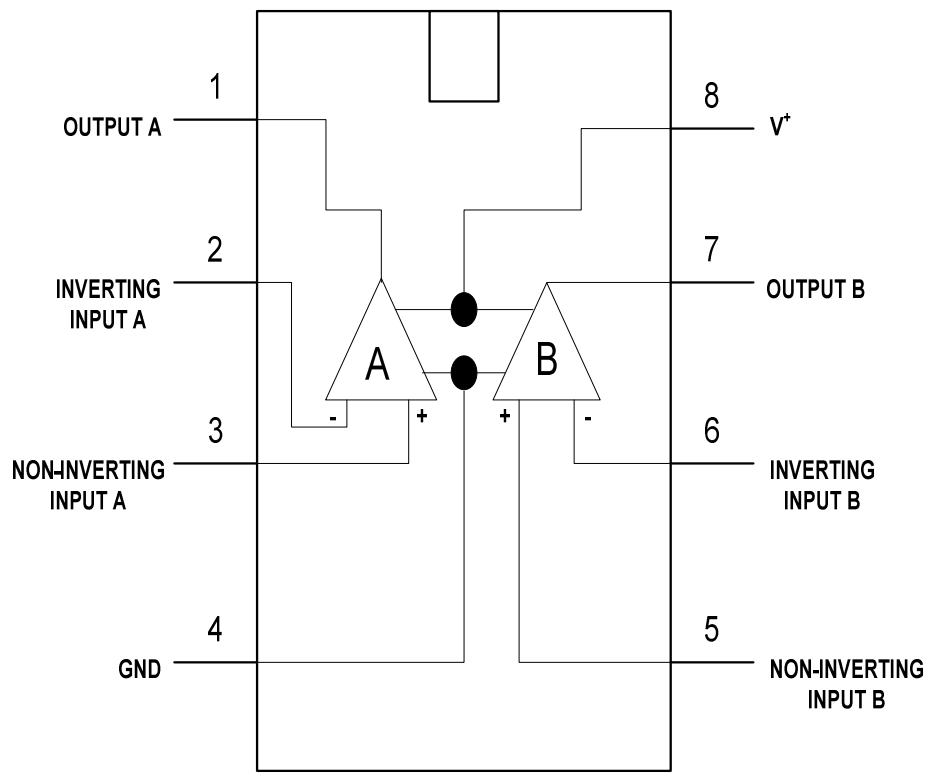

Fig. 7: Top view of LM393

### 3.1.6.3. LM324

The LM 324 series consists of four independent high gains and internally frequency compensated operational amplifiers which were designed specifically to operate from a single power supply over a wide range of voltages. Operation from split power supplies is also possible and the low power supply current drain is independent of the magnitude of the power supply voltage.

**3.1.6.4. IC TIMERS**

The emanation of IC timers eliminated a wide range of mechanical and electromechanical timing devices. It also helped in the generation of clock and oscillator circuits. Timing circuits are those, which will provide an output change after a predetermined time interval. This is of course, the action of the Monostable multivibrator which will give a time delay after a fraction of a second to several minutes quite accurately. The most popular of the present IC, which is available in an eight-pin dual inline package in both bipolar and CMOS form is 555 Timer. The 555 timers are a relatively stable IC capable of being operated as an accurate bi-stable, Monostable or Astable multivibrator. The timer comprises of 23 transistors, 2 diodes and 16 resistors in its internal circuitry. Its functional diagram is shown in Fig. 8.

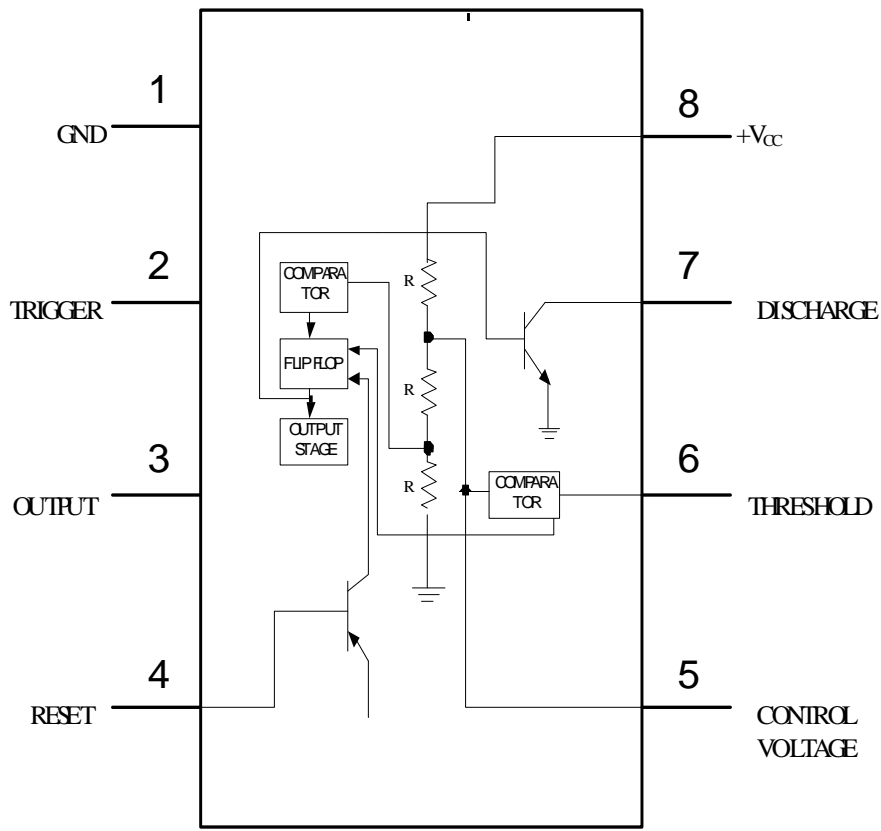

Fig. 8: Functional diagram of 555 Timer

The functional diagram consists of two comparators, a flip-flop, two control transistors and a high current output stage. Two comparators are actually operational amplifiers that compare input voltage to an internal reference voltage which are generated by an internal voltage divider of three 5K resistors. The reference voltages provided are one third and two third of $V_{CC}$. When the input voltage to either of the comparators is higher than the reference voltage for the comparator, the amplifier goes to saturation and produces an output signal to trigger the flip-flop. The output of the flip-flop controls the output stage of the timer. The 555 timer chip works from a D.C supply between 3-15V and can source or sink up to 200mA at its output. The operation of the 555 Timer is further defining the functions of all pins. The details regarding connection to be made to pins are as follows.

**Pin 1**: This is ground pin and should be connected to the negative side of the power supply voltage.

**Pin 2**: This is the trigger input. A negative going voltage pulse applied to this pin when falling below 1/3 $V_{CC}$ causes the comparator output to change state. The output level then switches from LOW to HIGH. The trigger pulse must be of shorter duration than the time interval set by the external CR network otherwise the output remains high until trigger input is driven high again.

**Pin 3**: This is the output pin and is capable of sinking or sourcing a load requiring up to 200mV and can drive TTL circuits. The output voltage available is approximately -1.7V

**Pin 4**: This is the rest pin and is used to reset the flip-flop that controls the state of output pin3. A reset is activated with a voltage level of between 0V and 0.4V and forces the output to go low regardless of the state of the other flip-flop inputs. If reset is not required, then pin 4 should be connected to the same point as pin 8 to prevent constant reset.

**Pin 5**: This is the control voltage input. A voltage applied to this point to this pin allows the timing variations independently of the external timing network. Control voltage may be varied from between 45 to 90 of the $V_{CC}$ value in Monostable mode. In Astable mode the variation is from 1.7 to the full value of supply voltage. This pin is connected to the internal voltage divider so that the voltage measurement from here to ground should read 2/3 of the voltage applied to pin 8. If this pin is not used it should be bypassed to ground, typically use a 10nF capacitor. This helps to maintain immunity from noise. The CMOS ICs for most applications will require the controlled voltage to be decoupled and it should be left unconnected.

**Pin 6**: This is the threshold input. It resets the flip-flop and hence drives the output low if the applied voltage rises above two-third of the voltage applied to pin 8. Additionally a current of minimum value 0.1 A must be supplied to this pin since this determines the maximum value of resistance that can be connected between the positive side of the supply and this pin. For a 15V supply the maximum value of resistance is 20M.

**Pin 7**: This is the discharge pin. It is connected to the collector of an NPN transistor while the emitter is grounded. Thus the transistor is turned on and pin 7 is effectively grounded. Usually the external timing capacitor is connected between pin 7 and ground and is thus discharged when the transistor goes on.

**Pin 8**: This is the power supply pin and is connected to the positive end of the supply voltage. The voltage applied may vary from 4.5V to 16V. But there are devices that operate up to 18V.

The reset input (555 pin 4) overrides all other inputs and the timing may be cancelled at any time by connecting reset to 0V, this instantly makes the output low and discharges the capacitor. If the reset function is not required the reset pin should he connected to +Vs.

### 3.1.6.4. Bistable Operation

The circuit is called a bistable because it is stable in two states: output high and output low. This is also known as a 'flip-flop'. It has two inputs: (1) Trigger (555 pin 2) makes the output high. The trigger is 'active low', it functions when $<1/3$ Vs. (2) Reset (555 pin 4) makes the output low. A reset is 'active low', it resets when $< 0.7V$

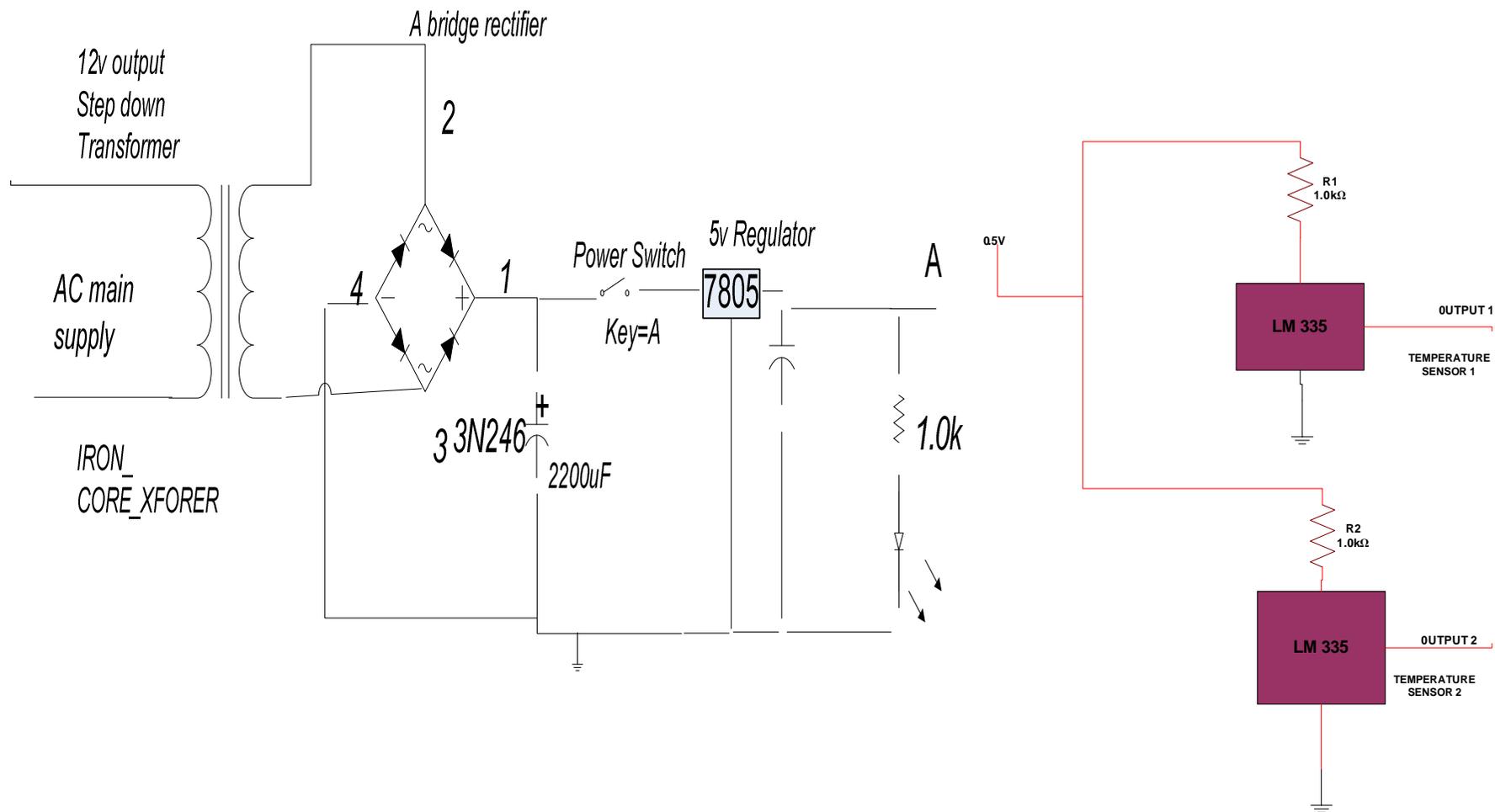

Fig. 9: Complete Circuit Diagram of a Dual Sensor Heat Control System

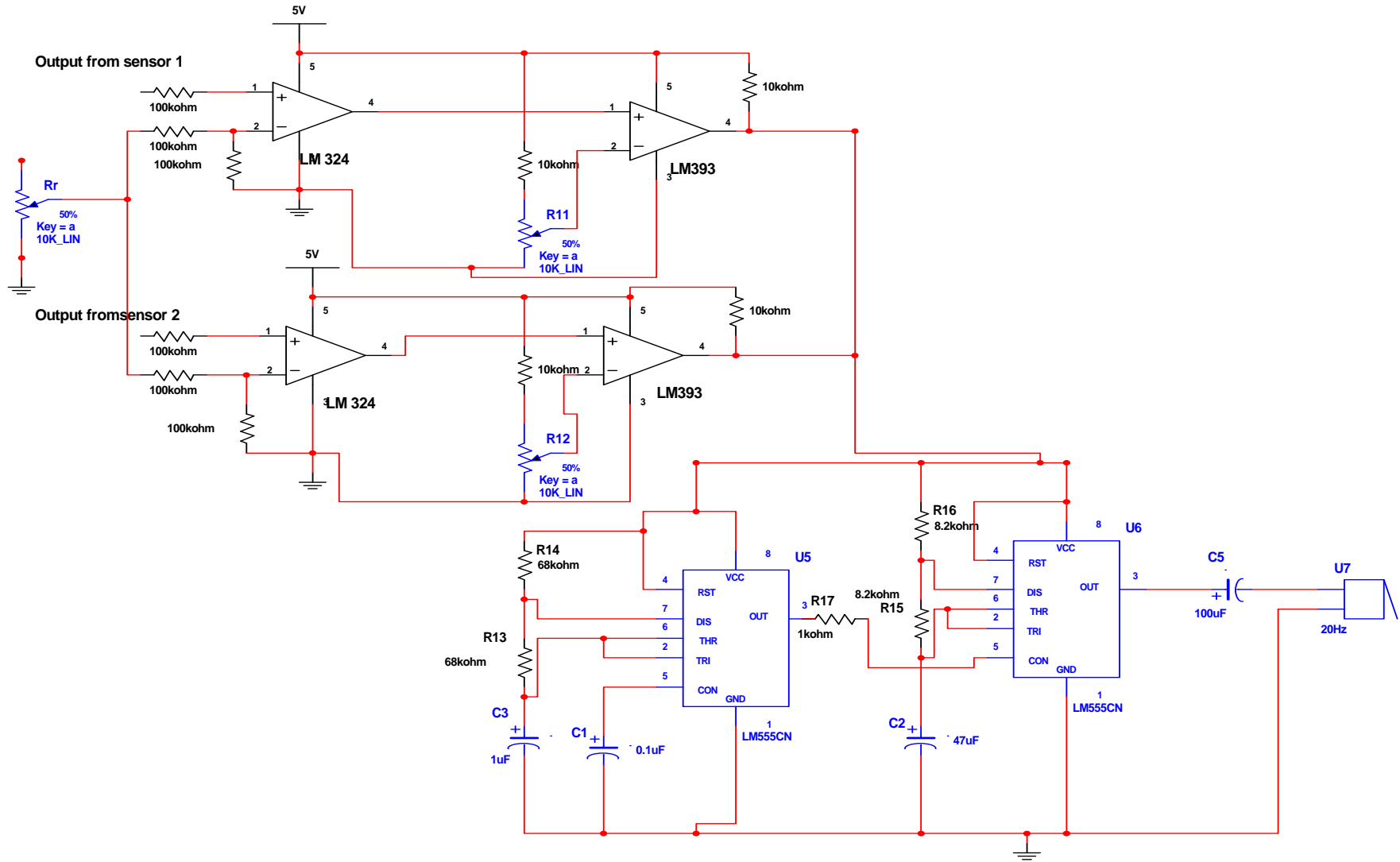

Fig. 9: Complete Circuit Diagram of a Dual Sensor Heat Control System (Cont'd)

# 4. CONSTRUCTION, TESTING AND DISCUSSION OF RESULTS

## 4.1. Circuit Construction

The circuit's construction involves making the circuit diagram on paper into reality. The first step involved mounting the circuit on the breadboard having temporary connection. The aim was to check the working ability of the circuit design before the permanent soldering of components on the Vero-board. The breadboard construction involved a close study of the circuit diagram. The success of the bread board construction led to the permanent connections (soldering) of components on the Vero-board in accordance with the circuit diagram. The circuit was divided into modules as demonstrated during the circuit design. Each module was constructed independently before they were joined together as one. The success of the bread board construction made it possible to start the construction on the Vero-board. It was a permanent connection of the components in accordance with the circuit diagram. The connection was achieved through soldering. The Vero-board was cut to suit the size of the circuit and it was further divided into regions consisting of the circuit's modules. Each module was completed independently and then connected together as a complete circuit. The power supply circuit which was quite sensitive was constructed with great care. Proper care was carried out to prevent or remove short circuits and power bridges.

The materials used during construction are: (1) Digital Multimeter, (2) Soldering iron, (3) cutting knife, (4) Scissors and (5) Pliers. While the materials used for the real construction are: (1) soldering lead, (2) Vero-board and (3) connecting wires.

## 4.1 CASING CONSTRUCTION

The casing was made of white rubber casing for system wiring. The full material was cut into size to fit the circuit. The rubber casing was chosen to enable ease of handling while making holes, screwing, and to serve as insulation for the circuit. The tools used for the construction of the casing include; a soldering iron for punching holes through the casing and a screwdriver was used for bolting the parts together.

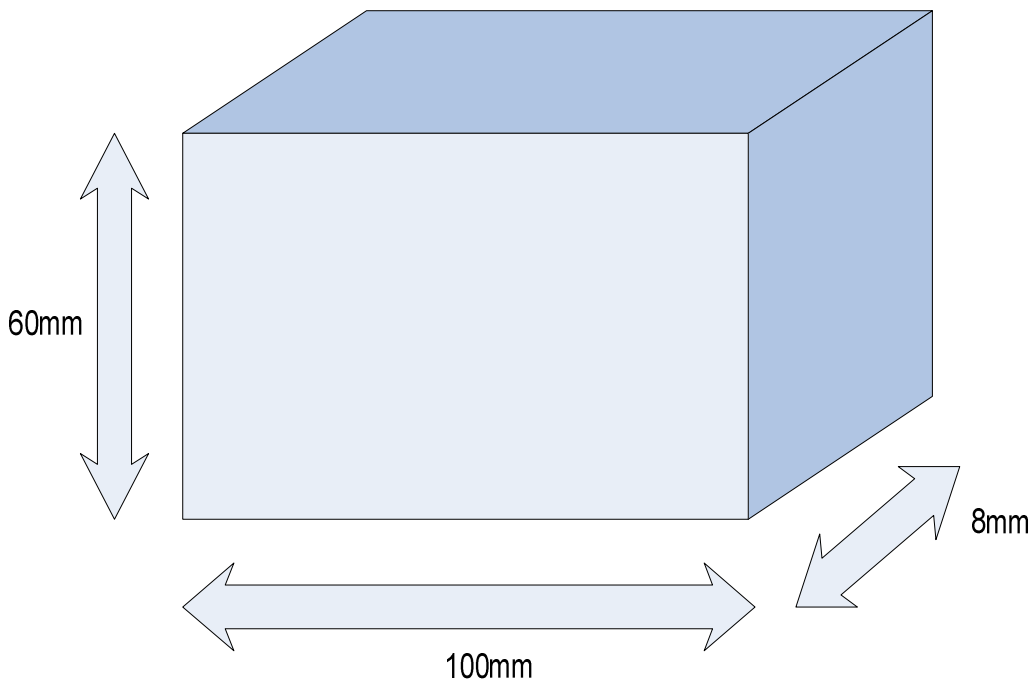

Fig. 10: Casing Dimension

### 4.2. Discussion of Results

The result was based on the fact that a corresponding alarm sound came 'ON' for a particular temperature sensor that has exceeded the set temperature level. The alarm was 'ON' during the time the input temperature was above the preset one i.e. The alarm indicates that a preset temperature has been exceeded. The fact that the involved temperature sensor (LM 335) was already calibrated during construction by the manufacturers increased the standard of the design and also eliminated the need to calibrate temperature while testing.

### 4.3. Testing

The main reason for testing all the components before they were finally soldered on the Vero-board is to avoid the painstaking effort it will take to dis-solder faulty components at the end of the day. From the continuity test carried out on the Vero-board to check the circuit path, it was discovered that the circuit was in a perfect working condition as continuity was ensured. Simulation of the circuit design was also done with the sole objective of comparing the results obtained from design calculations to that obtained from simulation. The two results when compared closely correspond with only a very slight discrepancy in values. The system testing was also carried out by adjusting the preset temperature of each temperature sensor to different values and observing the response of the device. Random values were selected on the preset temperature dial. Each temperature sensor was tested independently, and then both were tested together. The test was carried out several times to check the reliability of the device.

### 4.4. Problem Encountered

It should be noted at this juncture, that the realization of the final system work was not without problems. The various problems encountered during the design and implementation stage are highlighted as: (1) some basic components to be used for the system were not within reach as it was not available in town. (2) Some measuring instruments that would have been used for detailed analysis of the circuit (i.e. Oscilloscope, Transistor Tester) were limited for use. Simulation software was instead used for the circuit analysis. (3) The biggest problem encountered was on the implementation stage as the circuit alarm was triggered itself without temperature difference. The sensitivity of the circuit was reduced by reducing resistance R17 and making sure that there is no possibility of a short circuit to solve this problem.

## 5. CONCLUSIONS

It can be concluded that the sole aim of carrying out the design, analysis and implementation of *a dual sensor heat control system* was achieved, in that the aim was to develop a cheap, affordable, reliable and efficient temperature monitoring device, which was successfully realized at the end of the design process. One factor that accounts for the cheapness of the product was the proper choice of components used. The ones that were readily available were used, which a close substitute was found for those that were not readily available. The system involves the design, construction and testing of a heat monitoring device. It involves the triggering 'ON' of an alarm whenever the temperature of either of the devices being monitored exceeds a preset level. A temperature sensor (LM 335) was used to monitor variations in temperature level of the two devices to which it is connected. At the end of the construction, the work was tested to find out if the objectives of the research were met and it was discovered that the research functioned properly according to design. In the future, we intend to add a digital temperature display into the device to show the involved temperatures, use a wider range Temperature sensor to have more industrial importance, and interface the system with a computer to provide better control and monitoring.

## ACKNOWLEDGEMENTS

The authors would like to thank Col. Muhammed Sani Bello (RTD), OON, Vice Chairman of MTN Nigeria Communications Limited for supporting the research.

**Author**

**Engr. Adamu Murtala Zungeru** received his B.Eng. Degree in Electrical and Computer Engineering from the Federal University of Technology (FUT) Minna, Nigeria in 2004, and M.Sc. Degree in Electronic and Telecommunication Engineering from the Ahmadu Bello University (ABU) Zaria, Nigeria in 2009. He is a Lecturer Two (LII) at the Federal University of Technology Minna, Nigeria in 2005-till date. He is a registered Engineer with the Council for the Regulation of Engineering in Nigeria (COREN), Member of the Institute of Electrical and Electronics Engineers (IEEE), and a professional Member of the Association for Computing Machinery (ACM). He is currently a PhD candidate in the department of Electrical and Electronic Engineering at the University of Nottingham. His research interests are in the fields of automation, Security, swarm intelligence, routing, wireless sensor networks, energy harvesting, and energy management.

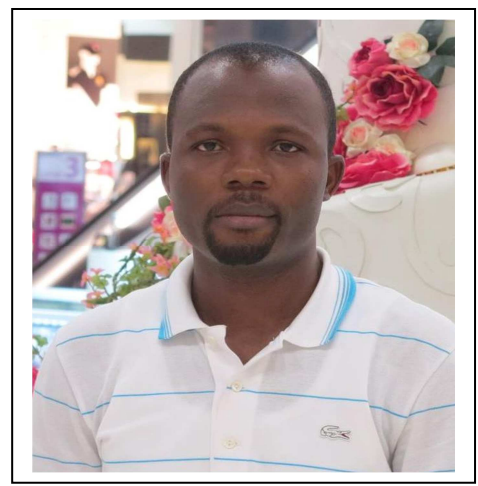